\documentclass[fleqn,10pt]{wlscirep}
\usepackage{graphicx,color}

\title{Export dynamics as an optimal growth problem in the network of global economy}

\author[1]{Michele Caraglio}
\author[1,*]{Fulvio Baldovin}
\author[1]{Attilio L. Stella}

\affil[1]{Dipartimento di Fisica e Astronomia and sezione INFN,
Universit\`a di Padova, Via Marzolo 8, I-35131 Padova, Italy}

\affil[*]{baldovin@pd.infn.it}

\keywords{complex systems, stochastic modeling, economic growth, cooperative effects}

\begin{abstract}
We analyze export data aggregated at world global level
of 219 classes of products over a period of 39 years. 
Our main goal is to set up a dynamical model to identify and
quantify plausible mechanisms by 
which the evolutions of the various exports affect each other.
This is pursued through a stochastic differential description,
partly inspired by approaches used in population 
dynamics or directed polymers in random media.
We outline a complex network of transfer rates which describes how resources are shifted
between different product classes, and determines how casual favorable
conditions for one export can spread to the other ones.
A calibration procedure allows to fit four free
model-parameters such that the dynamical evolution becomes consistent with
the average growth, the fluctuations, and the ranking of the export values 
observed in real data. 
Growth crucially depends on the balance between maintaining
and shifting resources to different exports, like in an explore-exploit
problem. Remarkably, the calibrated parameters warrant a close-to-maximum growth rate 
under the transient conditions realized in the period covered by data, 
implying an optimal self organization of the global export.
According to the model, major structural changes in the global economy 
take tens of years.
\end{abstract}

\begin{document}

\flushbottom
\maketitle
%
%
\thispagestyle{empty}

\section*{Introduction}

\noindent Issues concerning growth are of central importance 
in economic theories. Standard tools for analyzing economies evolution 
are empirical regressions, which allow to extract trends,
estimate fluctuations and produce forecasts. 
When data are referring
to a multitude of different productions, investments, or countries, 
one faces the problem of understanding how the relations among 
a large number of entities influence the overall dynamics. 
The problem then does not reduce
to the estimation of individual or average trends, but also involves the 
identification and quantification of collective effects possibly
influencing such trends and fluctuations. Through the estimation of
these effects we learn something about the
organization and functioning of the economy. In particular, we can
establish by which mechanisms and up to what extent the collective
effects contribute to the overall growth.

Such goals are pursued here by the construction of a minimal model
describing how the values of various global exports evolve in time. 
Theoretical modeling based on stochastic differential equations is
widely applied since long to different 
problems with growth-related features, like population and
evolutionary dynamics~\cite{nelson}, portfolio strategies~\cite{bouchaud_potters}, 
interface growth \cite{barabasi_stanley}
or optimal pinning of vortices by random defects in materials~\cite{zhang}. 
In economic settings, when for example representing the dynamics of
activities in various sectors,  $i$, 
one of the aspects to be modeled
concerns the alternance of favorable and unfavorable
conditions for growth of the sector's aggregated value $Z_i$, due to raw material prices,
innovations, etc.\,. This can be done
by introducing a multiplicative random noise term, $\eta_i\;Z_i$, 
in the expression of the time derivative of $Z_i$, with $\eta_i$ representing a 
Gaussian noise with zero average. 
Since economical trends are characterized by a typical duration $\tau$, 
the noise terms must preserve some memory of the past
values and thus be correlated in time. 
In addition, the representation of an intertwined economy demands for
the existence of proper links connecting different sectors. Hence, the
dynamical modeling must also include coupling parameters $J_{ij}$ describing
the shift of resources from production $j$ to production $i$.
Following Ref.~\cite{bouchaud_prl}, one then may write
\begin{equation}
\dfrac{\partial Z_i(t)}{\partial t} =
\sum_{j \neq i} \left[J_{ij} Z_j(t) -J_{ji} Z_i(t) \right] + 
\eta_i(t)\;Z_i(t), 
\label{eq_bouchaud_0}
\end{equation}
with
\begin{equation}
\langle\eta_i(t_1)\;\eta_i(t_2)\rangle
\propto\frac{1}{\tau}\;\mathrm{e}^{-|t_1-t_2|/\tau}
\end{equation}
(in the limit $\tau\to0$,
Brownian noise, $\delta$-correlated in time, is recovered). 
Interestingly, the combination of 
noise $\eta_i$ correlated in time and coupling terms $J_{ij}$ 
may induce a nonzero average growth even in the absence
of deterministic trends driving the single productions~\cite{bouchaud_prl}.
Thus, a most relevant question arises of
identifying the optimal coupling conditions which maximize the average
growth for a given noise level.
These conditions should realize the best compromise between 
maintaining investments in the same sector or shifting them to
other ones, providing an instance of the exploration-exploitation 
tradeoff problem~\cite{bouchaud_prl,cohen,march}.
Previous studies of this problem have been restricted to 
simplified situations, where the $Z_i$ play
equivalent roles and the noise terms, albeit correlated in time,
are not cross-influencing each other. 
Moreover, the conditions of the explore-exploit
dilemma have been discussed for asymptotically long times only~\cite{bouchaud_prl}. 
Our ambition here is to set up a sensible model adequate to address
the complexity of real economic data and to investigate how the
exploration-exploitation issue is possibly posed at the global economy
level and at accessible, finite times.

Network structures 
are well established tools~\cite{Vespignani_book} 
to address economic
complexity, especially for growth-related 
issues~\cite{garlaschelli,stanley_complex_net,barabasi,hidalgo,pietronero_1,pietronero_2}. 
Among the goals of the network approach is that of identifying
by data-analysis factors influencing the growth potential
of a given country, or that of explaining why, in spite of globalization,
certain countries are not able to develop productions which would substantially 
increase their Gross Domestic Product. Within these contexts, exchanges
among countries can be represented by a network (World Trade Network) in which
countries themselves are the nodes, and links are weighted 
by the existing amount of trade~\cite{garlaschelli,chakraborti}.
Another possibility is that of considering the exchanged products 
as nodes, while the strength of the links represents, e.g., the
proximity of pairs of products they connect -- a measure of the capability of a generic
country to produce them simultaneously at a significant level~\cite{barabasi}.
Based on such networks various schemes for assessing the fitness and the 
growth prospects of different countries have been recently
proposed~\cite{barabasi,hidalgo,pietronero_1,pietronero_2}.
However, dynamical considerations developed within these   
approaches do not rely on the formulation of an autonomous dynamical model
with explicit time dependence, as proposed here. 

A key feature revealed by our empirical analysis is that,
apart from few exceptions, on average the set of exported products
displays in time a stable ranking in terms of monetary values.
This circumstance enables us to identify a nontrivial network
structure which appropriately rules the dynamics.
Based on our model we are able to infer from the data 
basic features of the observed exports dynamics, which include 
the ingredients of an explore-exploit problem.
Besides quantifying the 
contribution to the growth determined by the coupling between 
products, we characterize, in an average sense, the time-scale 
and the extent to which favorable and unfavorable conditions 
influence each production. Useful insights are also provided
about the global response to changes of economic
fundamentals redefining the exchange network structure.

\section*{Export-value dynamics}
We use international trade data furnished by
the National Bureau of Economic Research~\cite{worldtradeflows} 
to extract yearly exports of 219 product classes
on the basis of the Standardized International Trade Code at
3-digit level (SITC-3), in a period of 39 years from 1962 to 2000. This
database is corrected for discrepancies among records provided by
the exporting and importing countries. 
The exports are reported in terms of their value in US-dollars. 
In this respect, one of the factors to be taken 
into account in estimating growth is the contribution due to the rate of
inflation/deflation of the dollar in each year (see below).

We consider for each product class the aggregate export
realized by all countries in one year.
Let us call $Z_{i,n}$ the total value of product $i$ 
($i=1,2,\dots,219$) exported in the year $n$ ($n=0,1,2,\dots,38$).
Figure~1 reports these quantities
as a function of $n$. The entries for each $i$ are
reported with a color whose wavelength is proportional
to the fraction $z_i$ of the value of product $i$ over the global yearly export,
averaged on the last 10 years covered by the database:
\begin{equation}
\label{eq_z_i}
z_i\equiv\frac{1}{10}\sum_{n=29}^{38}\frac{Z_{i,n}}{\sum_{k=1}^{219}Z_{k,n}} \; .
\end{equation}
It is evident in Figure~1 that the various lines, while indicating
an approximate average exponential growth, up to fluctuations and a few
exceptions do not cross and mix much over the whole period: a rainbow image is
perceived. One of the exceptions is represented, e.g., by electronic products,
which in the first part of the examined period display a rapid and 
substantial raise of value, which stabilizes then at a sensibly higher rank than the
initial one. Such an effect reflects the great technological advances after the
sixties and seventies. Similar transient behaviors occur for  
other products. This motivates why in Eq.~\eqref{eq_z_i} we focused on 
the last 10 years of the database, as better representative of a stabilizing
ranking.

\begin{figure}
\centerline{\includegraphics[width=.4\textwidth]{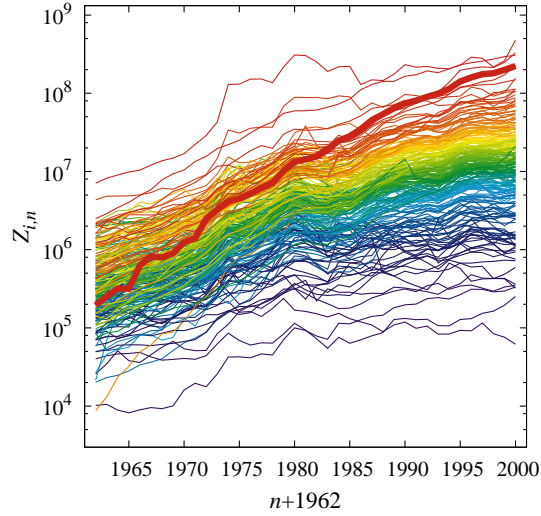}}
\caption{Time evolution of the yearly global value of exported products 
  from 1962 to 2000.
  Electronics goods are highlighted in thicker line.
}
\label{fig_ev_real}
\end{figure}

Switching to continuous time $t$ in year units (with $t=0$ corresponding to 1962) 
as a convenient choice for building up our model, 
we now indicate by $Z_i(t)$ the export-value   
of product $i$ in the year preceding $t$. 
$Z_{i,n}$ gives a discrete representation 
of this function in 39 points. The choice of continuous time 
is legitimated by the fact that yearly records are the result
of changes at much shorter time scales. 
Conceivably, in the simplest model one could think about,
$Z_i(t)$ would perform a sort of geometric Brownian motion, 
exponentially increasing at a rate $\mu$ (approximately common
to all $i$) and with fluctuations of amplitude $\sigma$ (volatility).
This would be consistent with the
fact that Figure~1 displays, for different goods $i$, 
akin fluctuations and a similar, almost constant average logarithmic slope.
However, trajectories of the various products would not
tend to the ranking shown by real data, and within such a model 
the rainbow effect would be progressively lost rather than being
progressively emphasized.
As anticipated in the Introduction,
mechanisms through which different exports influence 
each other
can be
specified 
by a matrix $\mathbf{J}=\left(J_{ij}\right)_{1\leq i,j\leq219}$
describing the transfer of value from export $j$ to export $i$.
So, we generalize Eq.~\eqref{eq_bouchaud_0} to
\begin{equation}
\dfrac{\partial Z_i(t)}{\partial t} =
\sum_{j \neq i} \left[J_{ij} Z_j(t) -J_{ji} Z_i(t) \right] + 
\left[\eta_i(t)+\mu(t)\right] Z_i(t) =
\sum_{j} A_{ij} Z_j(t) + 
\left[\eta_i(t)+\mu(t)\right] Z_i(t) \;,
\label{eq_bouchaud}
\end{equation}
where the rate matrix $\mathbf{A}$,  with elements 
$A_{ij}=J_{ij}$ for $j\neq i$ and $A_{ii}\equiv-\sum_{j\neq i} J_{ji}$,  
is analogous to the transition matrix defined for 
continuous-time Markov chains \cite{van_kampen}.
Here $\mu(t)$ represents a deterministic drift, accounting for the average
growth of the exports (including the inflationary one) in the absence of mutual
influences. $\eta_i(t)$ is a multiplicative noise source representing the variability 
of conditions faced by different products at different times. 
Eq.~\eqref{eq_bouchaud} is invariant with respect to changes of
the monetary unit for the $Z_i$-values \cite{mezard}.

For the purposes of the present work, in Eq.~\eqref{eq_bouchaud}
$\mathbf{J}$ does not exclusively represent direct shifts of
investments between different sectors of export. More generally, $\mathbf{J}$
also reflects transfers of resources which are necessary for the production
or the utilization of goods. For instance, the increase in the export of cars
is likely to determine an increase in the export of oil,
even if the two products do not represent a typical alternative
for investors and/or producers. As a consequence, the character of the
explore-exploit problem based on $\mathbf{J}$ does not
entirely depend on strategic investment policies~\cite{cohen,march}, 
but also reflects
intrinsic allocation constraints determined by the economy structure. 
In most applications (see, e.g.,~\cite{bouchaud_prl} and references
therein), the interaction matrix $\mathbf{J}$ is 
short-range and reflects the regularity of a lattice, so 
that the various $Z_i$'s turn out to be equivalent. 
In our case  different goods  
are evidently not equivalent, so $\mathbf{J}$ must play 
the important role of establishing and maintaining
the observed ranking of the various $Z_i$'s. 
Indeed, in the absence of noise terms the solutions of Eq.~\eqref{eq_bouchaud}   
tend to a long-time attractor in which the ranking of the $Z_i$'s
is determined by the kernel of $\mathbf{A}$ itself~\cite{rate_matrix}. 
We can exploit this feature by choosing 
\begin{equation}
\label{eq_j}
J_{ij} \propto z_i,
\end{equation}
since in this way $z_i$ itself would be in the kernel.
With the last choice, it is easy to see that on average the ranking approached
by the solutions of Eq.~\eqref{eq_bouchaud} is the one imposed by
$z_i$ itself and thus manifested by data. 
Eq.~\eqref{eq_j} has also the appeal of being
consistent with the gravity law, often used for estimating  
transfer rates in economics~\cite{grav_model_1,grav_model_2}: 
it is plausible for the shift of resources from export $j$ to export $i$ to be proportional
to $Z_i(t)\,Z_j(t)$, with $Z_i(t)$ and $Z_j(t)$ 
acting like masses in a gravity-force law.
The specific choice in Eq.~\eqref{eq_j}
accounts for such a proportionality
in a time-averaged sense. The gravity law
has an additional ingredient, an inverse proportionality to a 
power law of the ``distance'' between the products. 
In economic applications,
distance is often an elusive concept~\cite{grav_model_3}.
In our case, it should measure how far is a given product
from another one to the purpose of directly affecting its export.
One possibility is that of focusing on the empirical correlation
existing between the simultaneous variations of the exports of 
two products as a proxy to their inverse distance. 
Thus, we replace the notion of inverse-distance 
with the modulus $|c_{ij}|$ of the empirical correlator
of the logarithmic returns $R_{i,n}=\ln (Z_{i,n}/Z_{i,n-1})$:
\begin{equation}
c_{ij}\equiv
\frac{1}{38}
\;\sum_{n=1}^{38}r_{i,n}\;r_{j,n},
\label{eq_c_ij}
\end{equation}
where 
$r_{i,n}\equiv 
\frac{
R_{i,n}-
\sum_{n=1}^{38}R_{i,n}/38
}{
\sqrt{
\sum_{n=1}^{38}R_{i,n}^2/38
-(\sum_{n=1}^{38}R_{i,n}/38)^2
}
}
$
is the normalized return.
Notice that since Eq.~\eqref{eq_c_ij} involves the product of empirical returns at the same
year $n$, $c_{ij}$ carries no information about
time-correlations. It only encodes the cross-correlations of the products exports at the same time,
averaged over the whole period covered by the dataset. 
In summary, we thus write
\begin{equation}
\label{eq_Jij}
J_{ij}= G\;z_i\;|c_{ij}|\qquad(i\neq j),
\end{equation}
where $G$ is a coupling constant, and the factor $|c_{ij}|$
takes into account how much the variations of product $j$
influence those of product $i$.
The coupling matrix $\mathbf{J}$ defines
a weighted, oriented network connecting as nodes the various products.
The plausibility of the form adopted here in Eq.~\eqref{eq_Jij} has to
be ultimately tested by comparison with the data.

In Eq.~\eqref{eq_bouchaud} 
$\eta_i$ is a zero-average Gaussian noise term, describing growth or decay 
conjunctures for the product-value $Z_i$. These random occurrences are 
expected to be correlated both in time (in order to represent the existence 
of opportunity or crisis trends) and among products (since these trends
often affect simultaneously more than one export):
\begin{equation}
\langle\eta_i(t_1)\;\eta_j(t_2)\rangle
=d_{ij}\;\frac{\sigma^2}{\tau}\;\mathrm{e}^{-|t_1-t_2|/\tau}.
\end{equation}
The parameter $\tau$ is thus defining the typical opportunity or crisis 
duration, while $\sigma$ weights the importance of the stochastic part 
of the dynamics. In most growth models~\cite{nelson,zhang,bouchaud_prl,kpz} 
it is generally assumed $d_{ij}=\delta_{ij}$. 
We verified however that such an assumption
prevents Eq.~\eqref{eq_bouchaud} from quantitatively reproducing 
the correlator structure $c_{ij}$ detected empirically.
Instead, by choosing $d_{ij}=c_{ij}$,
the main features of the empirical correlators are
recovered to a satisfactory extent when simulating the calibrated model.

The term $\mu(t)$ in Eq.~\eqref{eq_bouchaud} is the deterministic
component of the growth rate, valid for each export independently of
the existence of the transfer network.
In principle, one should consider
deterministic trends to be both product-specific and time-dependent. 
Moreover, they should include an inflationary component beside the real one.
In favor of parameter parsimony, here we regard instead $\mu(t)$ as an
average growth rate $\overline\mu$ 
common to all products of the global economy, and we further restrict
its time-dependence to the inflationary component $I(t)$ alone:
\begin{equation}
\mu(t)=\overline\mu+I(t).
\end{equation}
Since $I(t)$ can be read as an yearly step-wise function  
from the Consumer Price Index reports~\cite{cpi_report} (see table~1),
in such a way we are thus left with the free parameter $\overline\mu$,
describing the network-independent 
average global growth rate over the observed time-span. 
One of our goals is its estimation. 
This is not trivial, as part of the 
empirical growth may stem from the capability of the network to spread 
opportunity trends among different exports.

\section*{Results and discussion}
As explained in the Methods section, a calibration procedure
applied over the dataset allows us to determine the four parameters
$G$, $\sigma$, $\tau$, $\overline\mu$, in ordered sequence.
Our first result is the reproduction of the basic qualitative
features displayed in Figure~1.
Taking into account that some characteristics of the dataset are 
embedded in the model definition (namely, the last 10-year ranking in $z_i$ and
the cross-correlation structure in $c_{ij}$), this result provides primarily a consistency test. 
However, the model itself offers a clue for 
the complexity of the data through four parameters and as such it
conveys novel information.
In particular, we will show below the existence of an optimal value
$G$ for maximum growth. 
Moreover, since the time correlation structure is not
{\it a priori} contained within the model, non-trivial insights about
the response properties follows.   

Not only the ``rainbow'' effect is replicated by the calibrated-model 
time evolution reported in Figure~2, but even 
specific exports (like electronic products) which in 1962 were out
of ranking, in the last years of
the evolution are attracted to a proper position.
A quantitative way of assessing such a convergence is
to consider the
Spearman's rank correlation coefficient $-1\leq r_s\leq 1$~\cite{spearman},
between results of the model time evolution and
the empirical $z_i$'s in Eq.~\eqref{eq_z_i}.
When $r_s=1$, the two compared rankings coincide; if $r_s=-1$,
their ordering is opposite.
Starting with the 1962's initial data, $Z_i(0)=Z_{i,0}$,
Figure~3 displays the time evolution of
the Spearman's rank correlator $r_s(t)$.
While for $G\neq0$ it is evident the role played by $J_{ij}$ in attracting the
initial data to the proper ranking,
the fact that $r_s(\infty)$ remains
smaller than one is due to the random fluctuations.
This ranking is also consistent with a very common stylized feature in Economy.   
Indeed, data analysis applied to both the model and the empirical
records reveal that the export value is distributed among
the various $Z_i$'s according to a Pareto-type power-law~\cite{mezard}.

\begin{figure}
\centerline{\includegraphics[width=.4\textwidth]{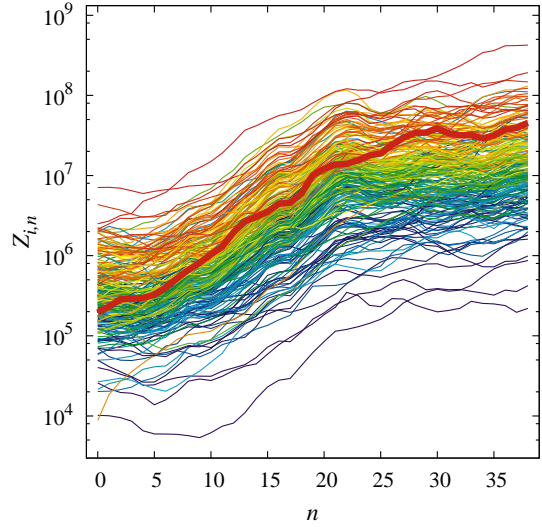}}
\caption{Model time-evolution of the export values starting from the
  real 1962 initial conditions. Color wavelength is assigned as in
  Figure~1. 
  Electronics goods are highlighted with a thicker line.
}\label{fig_ev_synthetic_self}
\end{figure}

The analysis of the Spearman coefficient
is also revealing in terms of the time-response to changes in the
economy fundamentals which may sensibly alter the product ranking 
(changes of the transfer rates $J_{ij}$ in our model). 
Indeed, the average behavior of $r_s(t)$ is well approximated
by the exponential law
\begin{equation}
\label{eq_exponential}
r_s(t)-r_s(\infty)
=
[r_s(0)-r_s(\infty)] 
\;\mathrm{e}^{-t/\tau_s},
\end{equation}
with a characteristic time, within our calibrated model, of
$\tau_s\simeq 41\;{\rm y}$ (see Figure~3). 
One could equivalently say that structural changes in the global
economy implying a replacement of the dominant products 
(like, e.g., those required to transform an oil-based economy into a green economy)
need tens of years to become effective.
In general, dynamical evolutions of our model starting from the 1962 historical
initial conditions show that $\tau_s$ is inversely proportional to the
transfer-rate parameter $G$, at fixed values of the other parameters.

Another result supporting the consistency of the model
concerns the cross-correlators $c_{ij}$ that one can reconstruct
from the numerical simulations. 
In Figure~4 we compare the correlators estimated
from the historical data of a
subset of nine exports (boxes below the diagonal) 
with those obtained by numerical
simulations of the calibrated model with historical initial conditions
(boxes above the diagonal).  
On purpose, three of the exports are chosen in the low, three in the
medium, and three in the high rank region.
Notice the high symmetry of the plot with respect to the diagonal.

\begin{figure}
\centerline{\includegraphics[width=.4\textwidth]{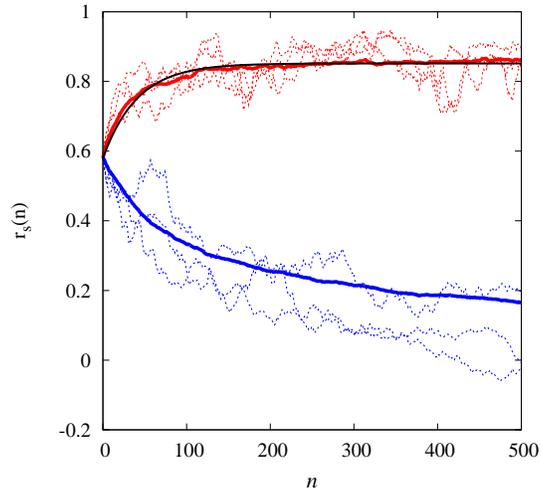}}
\caption{
  Time evolution of the Spearman's rank correlation coefficient.
  The initial condition $Z_i(0)=Z_{i,0}$ corresponds to the situation in 1962.
  Full red line: average over $100$ dynamical evolutions with
  calibrated parameters.
  Dotted lines refer to single dynamical realizations. 
  The full black line is a best exponential fit based on 
  Eq.~\eqref{eq_exponential} and yielding $\tau_s=41\;{\rm
  y}$.
  The parameter $G$ is set to zero for the blue lines.
}\label{fig_spearman}
\end{figure}

\begin{figure}
\centerline{\includegraphics[width=0.4\textwidth]{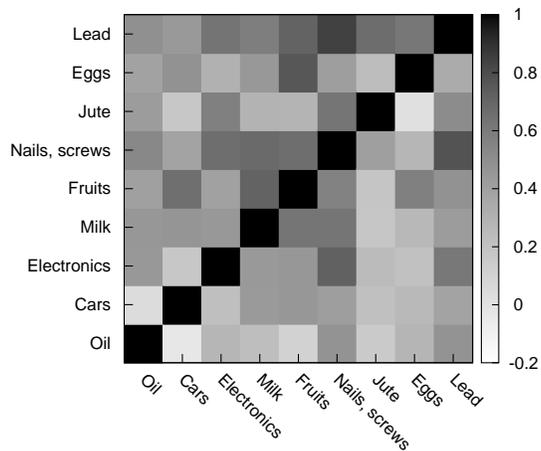}}
\caption{
  Comparison between historical correlators (above diagonal) and correlators 
  obtained by averaging over twenty histories generated from the calibrated model
  (below diagonal).
  The correlators pertain to 9 exports equally divided among low, medium and
  high ranking. A perfect agreement would imply a diagonal-symmetric
  plot.
}\label{fig_Correlators}
\end{figure}

An intriguing feature of the model in Eq.~\eqref{eq_bouchaud} 
is that not all of the observed growth comes from the deterministic rate 
$\mu(t)$.
Indeed, in view of the multiplicative nature of the terms in the equations,
local favorable stochastic fluctuations should spread out the
network more efficiently than negative random trends, enhancing the
average growth. This complies to the dilemma between exploiting a local
product's opportunity (of average intensity $\sigma$ and expected duration $\tau$)
and exploring other products' chances throughout the network by transferring 
part of the local value (at a rate controlled by $G$)~\cite{bouchaud_prl}.
Given the calibrated value for $\sigma$, $\tau$, and $\overline\mu$,
we studied the average growth rate over the time $T$,
\begin{equation}
\lambda_T\equiv\frac{1}{219\;T}\sum_{i=1}^{219}\ln\frac{Z_{i}(T)}{Z_{i}(0)},
\label{eq_lambda}
\end{equation}
as a function of the transfer coupling constant $G$.
If $T\lesssim\tau_s$, Eq.~\eqref{eq_lambda} defines 
what we call a transient growth rate;
it is only in the limit $T\gg\tau_s$ that $\lambda_T$ becomes the
asymptotic growth rate associated to a steady-state evolution.
In Figure~5 we report the curves corresponding to $\lambda_{T}$, with
$T=38\;{\rm y}$ (red) and $T\gg\tau_s$ (blue). 
As $G$ increases, $\lambda_T$ increases with respect to the average
deterministic growth rate and, after reaching a maximum, decreases. 
For large values of $G$
the transient growth rate becomes even smaller that the deterministic one. 
This effect, which does not show up for the steady-state growth rate,
is due to specificities of the initial conditions which imply 
some large transfers in the initial stages of the
dynamics.
Remarkably, the transient growth-rate attains its
maximum very close to the value of $G\simeq 0.05\;{\rm y^{-1}}$ obtained in our
calibration. 
Taking into account the weight of the links,
$\sum_{i,j\neq i}J_{ij}/N$,
this corresponds to an average yearly transfer of $2\%$ in value for
each product.
We thus conclude that according to our model the global economy
network appears to be
self-organized to guarantee close-to-maximum transient growth conditions.
On the other hand, on much longer times (corresponding to the
stationary regime) $G$
should be one order of magnitude larger than the calibrated one for
optimal growth conditions. 
Of course, the features of this
stationary regime depend on the assumption of the absence of
major changes in the structure of the economy; indeed, the latter could  
modify the link-weights $J_{ij}$.

\begin{figure}
\centerline{\includegraphics[width=.4\textwidth]{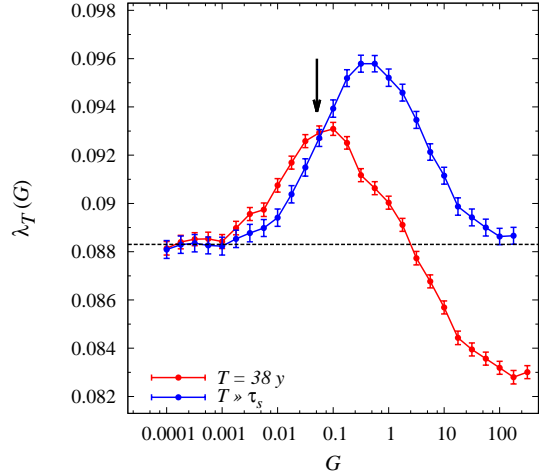}}
\caption{
  Average growth rate $\lambda_T$ as a function of $G$.
  The red line refers to $T=38\;{\rm y}$, the blue line to the
  asymptotic growth rate obtained with $T\gg\tau_s$. 
  The arrow indicates the calibrated value $G=0.051\pm0.005\;{\rm y^{-1}}$.
  The 
  dashed line corresponds to the average deterministic growth 
  $\overline\mu+\sum_{n=1}^{38}I(n)/38$.  
}\label{fig_growth}
\end{figure}

The above results descend from the structure of our model,
which explicitly singles out a deterministic component of the growth.
In addition, they crucially depend on the peculiar structure of our network
of transfer rates primarily dictated by the rank order. 
It is worth illustrating with an example the
specificity of this network. In Figure~6 we plot the weighted, oriented links among
vegetables, fruits, and oil, according to Eq.~\eqref{eq_Jij},
with heavier lines indicating larger transfer rates.
Contrary to naive expectation, 
the links of both vegetables and fruits to oil are
much stronger than those between vegetables and fruits
themselves. This emphasizes the need of value transferring from
agriculture goods to oil, involved for instance 
in a petrol-based agriculture production. 

\begin{figure}
\begin{center}
\setlength{\unitlength}{0.5 cm}
\begin{picture}(12,4.5)(0,0)
\put(0,0){\makebox(12,4.5){}}
\put(3,1){\circle*{0.4}}
\put(1,3.5){\circle*{0.4}}
\put(5,3.5){\circle*{0.4}}
\put(2.6,0.15){(o)}
\put(4.8,4.1){(f)}
\put(0.2,4.1){(v)}
\linethickness{0.40mm}
\qbezier(3,1)(1.7,2)(1,3.5)
\qbezier(1.73,2.1)(1.83,2.05)(1.93,2)
\qbezier(1.94,2.2)(1.94,2.1)(1.94,2)
\linethickness{0.30mm}
\qbezier(3,1)(4.3,2)(5,3.5)
\qbezier(4.07,2.2)(4.07,2.1)(4.07,2)
\qbezier(4.27,2.1)(4.17,2.05)(4.07,2)
\linethickness{0.20mm}
\qbezier(5,3.5)(3,3.9)(1,3.5)
\qbezier(2.9,3.8)(3,3.75)(3.1,3.7)
\qbezier(2.9,3.6)(3,3.65)(3.1,3.7)
\linethickness{0.15mm}
\qbezier(1,3.5)(3,3.1)(5,3.5)
\qbezier(3.1,3.4)(3,3.35)(2.9,3.3)
\qbezier(3.1,3.2)(3,3.25)(2.9,3.3)
\linethickness{0.05mm}
\qbezier(1,3.5)(2.3,2.5)(3,1)
\qbezier(2,2.34)(2,2.44)(2,2.54)
\qbezier(2.2,2.46)(2.1,2.5)(2,2.54)
\linethickness{0.05mm}
\qbezier(5,3.5)(3.7,2.5)(3,1)
\qbezier(4,2.35)(4,2.45)(4,2.55)
\qbezier(3.8,2.45)(3.9,2.5)(4,2.55)

\small
\put(7,3.2){$J_{of}=8\times10^{-3}\,G$}
\put(7,2.7){$J_{ov}=6\times10^{-3}\,G$}
\put(7,2.2){$J_{fo}=1\times10^{-3}\,G$}
\put(7,1.7){$J_{fv}=4\times10^{-3}\,G$}
\put(7,1.2){$J_{vo}=1\times10^{-3}\,G$}
\put(7,0.7){$J_{vf}=3\times10^{-3}\,G$}

\end{picture}
\end{center}
\medskip
\caption{Weighted links among vegetables (v), fruits (f), oil (o) goods
  according to Eq.~\eqref{eq_Jij}.
}
\label{fig_graph_vegetables}
\end{figure}
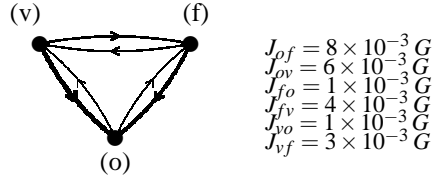

\section*{Conclusions}
We proposed an autonomous dynamical model describing the evolution of
the value of product exports within a global-economy complex network. 
Minimality has been an obliged feature in order to
highlight cooperative endogenous mechanisms underlying economic growth. 
Specifically, in our equations we singled out a deterministic growth term
including inflationary contributions, a stochastic one representing the
alternance of favorable and unfavorable conditions to the development
of each export class, and finally a network of value-transfers among
products. This last ingredient has been successfully identified primarily
through a ranking-based criterion. The stochastic component is characterized both by 
time-correlations associated to the duration of economic trends and by
product cross-correlations reflecting, e.g., the existence
of economy sectors in which several exports are simultaneously involved.
In spite of its parsimonious character, the model provides realistic
estimates of fluctuation properties, characteristic response times,
and average growth rates. It also provides an evaluation of the
average percentage of value transfer for the exports.
Importantly, through the distinction between $\overline\mu$ and the
empirical growth rate,  it characterizes which part of the growth
can be ascribed to transfer mechanisms determined by investments and
structural interdependences. 

We could verify that the global growth complies
with the typical conditions of an explore-exploit problem.
The optimal solution for such a problem 
depends on both structural interdependencies among products
and strategic investment choices.
Noteworthy, the calibrated network couplings
realize close-to-optimal conditions for maximal average growth in the
period covered by data. Thus, for the prevailing correlated noise conditions,
the network transfer rates appear to be self-organized towards 
a close to optimal solution of the explore-exploit dilemma.

We are confident that the results obtained for our model suggest
novel ways of analyzing and interpreting similar datasets
offering elements on which to base strategic choices.

\section*{Methods}
To solve Eq.~\eqref{eq_bouchaud}, one integrates the stochastic differential equations 
\begin{equation}
dZ_i(t)= \sum_{j \neq i} \left[J_{ij} Z_j(t) -J_{ji} Z_i(t) \right] dt+ 
\left[\eta_i(t)+\mu(t)\right] Z_i(t) \; dt\;,
\label{eq_bouchaud_sde}
\end{equation}
with the random noise $\eta_i(t)$ evolving according to
\begin{equation}
\eta_i(t)=\rho\,\eta_i(t-dt)+\sqrt{1-\rho^2}\;\dfrac{\sigma}{\sqrt{\tau}}\;\xi_i(t),
\end{equation}
where $\rho\equiv e^{-dt/\tau}$, and $\xi_i$ is Gaussian-distributed with zero mean and unit variance
satisfying $\langle\xi_{i}(t)\;\xi_j(t')\rangle=c_{ij}\,\delta(t-t')$.
Cross-correlation among the noise terms is attained by performing the
LDL decomposition~\cite{cholesky} 
(a variant of Cholesky decomposition) of 
the matrix \mbox{$C\equiv(c_{ij})_{i,j=1,\ldots,N}$}, i.e., $C=LDL^T$,
(see, e.g.,~\cite{cholesky}) and  applying it to a set of independent
Gaussians $\tilde\xi$ to generate $\xi=L D^{\frac{1}{2}}\, \tilde\xi$.
Under the natural assumption $dt\ll\tau$ different integration schemes (e.g., It\^o, Stratonovich)
provide the same results. 

The calibration of the four model parameters
$(G,\sigma,\tau,\overline\mu)$ proceeds as follows.
Eq.~\eqref{eq_bouchaud_sde} can be rewitten as 
\begin{equation}
d\ln Z_i(t)= \left\{\sum_{j \neq i} \left[J_{ij} \frac{Z_j(t)}{Z_i(t)}
  -J_{ji} \right]
+ \eta_i(t)+\mu(t)\right\} \; dt\;.
\label{eq_bouchaud_sde_ln}
\end{equation}
Integrating Eq.~\eqref{eq_bouchaud_sde_ln} between $n_1$ and $n_2$ we may write
\begin{equation}
\label{eq_f_i}
f_i(n_1,n_2)\simeq G\;g_i(n_1,n_2)+{\overline \mu}+\frac{1}{n_2-n_1}\int_{n_1}^{n_2}dt\;\eta_i(t),
\end{equation}
where $(n_2-n_1)f_i(n_1,n_2)\equiv\ln[Z_{i,n_2}/Z_{i,n_1}]-\int_{n_1}^{n_2}dt I(t)$ is provided by empirical data,
and
\begin{equation}
(n_2-n_1) g_i(n_1,n_2)\equiv\sum_{n=n_1+1}^{n_2}\sum_{j\neq i}
\left[
\frac{z_i\,|c_{ij}|}{2}
\left(
\frac{Z_{j,n-1}}{Z_{i,n-1}}
+
\frac{Z_{j,n}}{Z_{i,n}}
\right)
-z_j\,|c_{ij}|
\right]
\end{equation}
approximates through the 
empirical observations the stochastic integral involving the transfer terms.
The last term in Eq.~\eqref{eq_f_i} can be viewed as a source of random errors for $f_i$.
As a result, the parameter $G$ can be calibrated as the slope of a linear regression 
of $f_i$ vs $g_i$.
However, when $|g_i|$ is small the random source makes the
determination of $G$ not reliable. 
Through synthetic simulations of the model
with the same duration and initial conditions of the empirical data, we verified that
by considering in the above regression only 1/10 of the points with larger $|g_i|$, the calibration
procedure obtains the correct $G$-value within a confidence of 10\%. 
In Figure~7 we display the scatter plot of $f_i$ vs $g_i$,
indicating in blue the points selected for the linear regression
(dashed line).
In fact ${\overline \mu}$ should correspond to the intercept of the regression.
With our restriction to large $|g_i|$,
the estimate of $\overline{\mu}$ through this intercept is not very precise
and becomes sharper only if all $|g_i|$'s are considered. Below, we
choose an alternative method for the calibration of $\overline{\mu}$
(see below).

A straightforward calculation yields
\begin{equation}\label{eq_th_variance}
\mathbb E\left[\left(\int_{0}^{n}dt\;\eta(t)\right)^2\right]
-\mathbb E\left[\left(\int_{0}^{n}dt\;\eta(t)\right)\right]^2
=2 \, \sigma^2
\left[
n+\tau\left(
e^{-\frac{n}{\tau}}-1
\right)
\right] \, .
\end{equation}
We thus calibrate the parameters $\sigma$ and $\tau$ by fitting with
the r.h.s. of Eq.~\eqref{eq_th_variance}
the empirical variance of $n[f_i(0,n)-G\,g_i(0,n)]$, 
evaluated averaging over all products $i$ (see Figure~8).
We remind that such a variance is independent of $\mu(t)$. 

Finally, the parameter $\overline\mu$ is calibrated as the best fitting value
which, added to the inflation and to the growth due to the network,
numerically reproduces the overall growth of real exports in 
the time interval $[0,38]$ at given calibrated values for 
$G$, $\sigma$, $\tau$, starting from the historical initial conditions at $t=0$. 
The values $I(t)$ for the inflation rate are deduced from the CPI Detailed
Report made by Bureau of Labor Statistics~\cite{cpi_report} and
reported in  
table~1. 

The application of the above procedure gives in summary 
\begin{eqnarray}
G&=&0.051\pm0.005\;{\rm y}^{-1},\\
\sigma&=&0.098\pm0.005\;{\rm y}^{-1/2},\\
\tau&=&0.8^{+0.2}_{-0.7}\;{\rm y},\\
\overline\mu&=&0.041\pm 0.001\;{\rm y}^{-1}.
\end{eqnarray}
Parameter precisions are evaluated as standard deviations
resulting from repeated applications of the above calibration procedure 
to synthetic simulations of
Eq.~\eqref{eq_bouchaud_sde},
performed within the time interval $[0,38]$ at the calibrated value of the
parameters.

\begin{figure}
\centerline{\includegraphics[width=.3\textwidth]{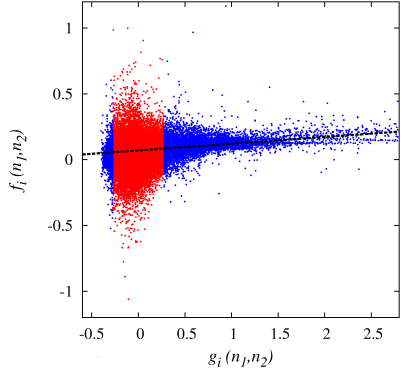}}
\caption{Calibration of the parameter $G$.
  Scatter plot of $f_i(n_1,n_2)$ vs. $g_i(n_1,n_2)$, with
  $i=1,\ldots,219$,
  $n_1=0,\ldots,37$, $n_2=n_1+1,\ldots,38$. 
  In blue (red) are the 10\% (90\%) of the points with larger (smaller) $|g_i|$.
  The dashed line is the linear
  regression $y=G\,x + k$, giving $G=0.051$ and $k=0.069$.}\label{fig_cal_J}
\end{figure}

\begin{figure}
\centerline{\includegraphics[width=.3\textwidth]{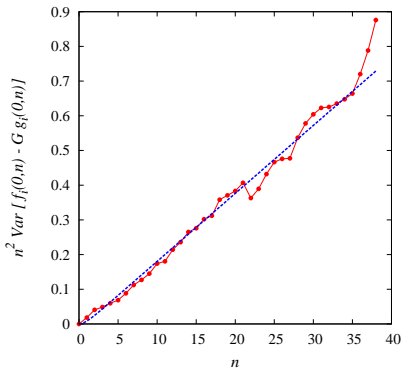}}
\caption{
  Calibration of the parameters $\sigma$ and $\tau$.
  Variance of $n[f_i(0,n)-G\,g_i(0,n)]$ as a function
  of $n$. Dashed line is a fit with the r.h.s. of 
  Eq.~\eqref{eq_th_variance}, giving 
  $\sigma=0.098$ and $\tau=0.8$.
}\label{fig_cal_ST}
\end{figure}

\begin{table}
\caption{Yearly inflation rate according to~\cite{cpi_report}. 
  The average value over the 38 years is $I=4.73$\%\\$\;$\\
}
\label{tab_inflation}
\begin{center}
  \begin{tabular}{|c|c||c|c||c|c||c|c||}
\hline year & $I_t$ & year & $I_t$ & year & $I_t$ & year & $I_t$ \\ 
\hline \hline 1963 & $1.3$ & 1964 & $1.3$ & 1965 & $1.6$ & 1966 & $2.9$ \\ 
\hline 1967 & $3.1$ & 1968 & $4.2$ & 1969 & $5.5$ & 1970 & $5.7$ \\ 
\hline 1971 & $4.4$ & 1972 & $3.2$ & 1973 & $6.2$ & 1974 & $11.0$ \\ 
\hline 1975 & $9.1$ & 1976 & $5.8$ & 1077 & $6.5$ & 1978 & $7.6$ \\ 
\hline 1979 & $11.3$ & 1980 & $13.5$ & 1981 & $10.3$ & 1982 & $6.2$ \\ 
\hline 1983 & $3.2$ & 1984 & $4.3$ & 1985 & $3.6$ & 1986 & $1.9$ \\ 
\hline 1987 & $3.6$ & 1988 & $4.1$ & 1989 & $4.8$ & 1990 & $5.4$ \\ 
\hline 1991 & $4.2$ & 1992 & $3.0$ & 1993 & $3.0$ & 1994 & $2.6$ \\ 
\hline 1995 & $2.8$ & 1996 & $3.0$ & 1997 & $2.3$ & 1998 & $1.6$ \\ 
\hline 1999 & $2.2$ & 2000 & $3.4$ &  &  &  &  \\ 
\hline 
\end{tabular}
\end{center}
\end{table}


\section*{Acknowledgments}
We acknowledge support from the JUNIOR Research Project ``Dynamical behavior of complex systems: from scaling symmetries to economic growth'' of the University of Padova.

\section*{Author contributions statement}
A.L.S. and F.B. conceived the research project. M.C. performed the research. All authors analyzed the results. All authors wrote the text of the manuscript and reviewed the manuscript. 

\section*{Additional information}
\paragraph*{Competing financial interests:} The authors declare no competing financial interests.

\end{document}